\documentclass[superscriptaddress,nofootinbib, amsmath,amssymb,preprintnumbers,prdfloatfix,twocolumn,PRL]{revtex4-2}
\usepackage{CJK}
\usepackage{graphicx}
\usepackage{dcolumn}
\usepackage{upgreek}
\usepackage{amsmath}
\usepackage{bm}
\usepackage[colorlinks=true,pdfstartview=FitV,breaklinks=true]{hyperref}
\usepackage[dvipsnames,table]{xcolor}
\hypersetup{urlcolor=BlueViolet,
	    citecolor=Plum,
	    linkcolor=PineGreen}
\usepackage{lipsum}	    
\usepackage{titlesec}
\usepackage{etoolbox}
\usepackage[normalem]{ulem}
\usepackage{tabularx}
\usepackage{amssymb}
\usepackage{float}	
\usepackage{caption}
\usepackage{subcaption}
\usepackage{comment}
\usepackage{appendix}
\usepackage{aasmacros}
\usepackage{subcaption}
\usepackage{afterpage}

\definecolor{offblue}{RGB}{23,80,153}

\newcommand{\htwo}{\mathrm{H_2}}
\newcommand{\hminus}{\mathrm{H^-}}
\newcommand{\h}{\mathrm{H}}

\newcommand{\esh}{\varepsilon_{sh}}

\newcommand{\equaref}[1]{Eq.~(\ref{#1})}

\newcommand{\figref}[1]{Fig.~\ref{#1}}

\newcommand{\refref}[1]{Ref.~\cite{#1}}

\newcommand{\ALP}{\textsc{ALP}}
\newcommand{\ALPs}{\textsc{ALP}s}

\begin{document}

\preprint{IPMU24-0013}

\title{Direct-collapse supermassive black holes from relic particle decay}

 \author{Yifan Lu}
 \email{yifanlu@g.ucla.edu}

 \author{Zachary S. C. Picker}
 \email{zpicker@physics.ucla.edu}

\affiliation{Department of Physics and Astronomy, University of California Los Angeles,\\ Los Angeles, California, 90095-1547, USA}

\author{Alexander Kusenko}
\email{kusenko@ucla.edu}

\affiliation{Department of Physics and Astronomy, University of California Los Angeles,\\ Los Angeles, California, 90095-1547, USA}
\affiliation{Kavli Institute for the Physics and Mathematics of the Universe (WPI), The University of Tokyo Institutes for Advanced Study, The University of Tokyo, Chiba 277-8583, Japan}

\begin{abstract}
\noindent 
We investigate the formation of high-redshift supermassive black holes (SMBHs) via the direct collapse of baryonic clouds, where the unwanted formation of molecular hydrogen is successfully suppressed by a Lyman-Werner (LW) photon background from relic particle decay. We improve on existing studies by dynamically simulating the collapse, accounting for the adiabatic contraction of the DM halo, as well as the \textit{in-situ} production of the LW photons within the cloud which reduce the impact of the cloud's shielding. We find a viable parameter space where the decay of either some of the dark matter or all of a subdominant decaying species successfully allows direct collapse of the cloud to a SMBH.
\end{abstract}

\maketitle

\noindent The formation of the supermassive black holes (SMBHs) which reside in active galactic nuclei (AGN) and distant quasars remains an open  question~\cite{Inayoshi:2019fun, mayer2018route, volonteri2010formation, wiklind2012first,Bogdan:2023ilu,sethi2010supermassive}, dramatically reignited by recent observations from the James Webb Space Telescope (JWST)~\cite{JWST2006SSRv..123..485G} of high-redshift \mbox{($z\gtrsim 6$)} active galactic nuclei~\cite{jwst2023A&A...677A.145U,jwst2023ApJ...953L..29L,jwst2023ApJ...959...39H,jwst2023Natur.619..716C,jwst2023ApJ...942L..17O,jwst2023ApJ...954L...4K,jwst2023ARA&A..61..373F,jwstMaiolino:2023zdu}. In light of these discoveries, many  proposals were made for the creation of these SMBHs, including from the death of Pop~III stars~\cite{Banik:2016qww,2009ApJ...706.1184O, Kroupa:2020sru}, gravothermal collapse of self-interacting dark matter halos~\cite{Balberg:2001qg, Pollack:2014rja, koda2011gravothermal, Feng:2020kxv, Xiao:2021ftk}, primordial black holes (PBHs)~\cite{pbh,Hawking:1971ei,Carr:1974nx,Chapline:1975ojl,Carr:2020gox,Bean:2002kx,Duechting:2004dk,Khlopov:2004sc,Dokuchaev:2004kr,Kawasaki:2012kn,Carr:2023tpt, Flores:2020drq, Flores:2021jas, Flores:2023zpf}, and finally the direct collapse of gas clouds~\cite{Inayoshi:2019fun,Bromm:2002hb, Choi:2013kia, 2010MNRAS.402.1249S,Biermann:2006bu,Stasielak:2006br,Spolyar:2007qv, Natarajan:2008db, Araya:2013dwa,Friedlander:2022ovf,dijkstra2008MNRAS.391.1961D,sethi2010supermassive,Lu:2023xoi}.

The latter `direct collapse' scenario succeeds if the gas collapses and forms one central black hole instead of fragmenting into disjoint smaller clouds. Cooling by molecular hydrogen ($\htwo$) leads to this fragmentation~\cite{lepp1983kinetic, hollenbach1979molecule} and therefore suppression of $\htwo$ formation ensures the direct collapse~\cite{Bromm:2002hb, Choi:2013kia, 2010MNRAS.402.1249S, 2010MNRAS.402.1249S, Oh:2001ex, Omukai:2008wv, Omukai:2000ic, becerra2018assembly, Jeon:2024iml}. This can be achieved either by direct dissociation~\cite{2010MNRAS.402.1249S,Haiman:1996rc,dijkstra2008MNRAS.391.1961D,Inayoshi:2019fun,Biermann:2006bu,Stasielak:2006br,Spolyar:2007qv, Natarajan:2008db, Araya:2013dwa,Friedlander:2022ovf,Cyr:2022urs,agarwal2012ubiquitous, dijkstra2014feedback, Shang:2009ij, Hartwig:2015ura, Inayoshi:2014nha, Latif:2014cda} or by excess heating~\cite{sethi2010supermassive,Lu:2023xoi}. The formation of a supermassive black hole from this successful collapse, including the resolution of the angular momentum transport problem \cite{Begelman:2006db,Haemmerle:2021tcj,Latif:2013dua,10.1093/mnras/stv694,2009ApJ...702L...5B,Inayoshi:2019fun,Shlosman:1989te, shlosman1990fuelling,Eisenstein:1994nh, koushiappas2004massive,Inayoshi:2014rda, VanBorm:2014vva, Becerra:2014xea, Regan:2013ppa,Chandrasekhar:1964zz, zel2014stars, Shapiro:1983du} and the formation and collapse of a supermassive star as an intermediate phase~\cite{Inayoshi:2014rda, VanBorm:2014vva, Becerra:2014xea, Regan:2013ppa, Chandrasekhar:1964zz, zel2014stars, Shapiro:1983du}, has been studied extensively and is now well-established.

It is natural to ask if either the excess heating or the required Lyman-Werner (LW) background for dissociation could be provided by the decay of a relatively light particle. This particle could either make up all of the dark matter (DM), in which case the decay rate must be small on the cosmological time scales, or it could be a subdominant component of DM which decays entirely around the epoch of reionization. 
Notably, the abundance of SMBH produced via this mechanism can be adequately explained by the expected mass function of halos at early times~\cite{Friedlander:2022ovf}.\\

In this {\it letter}, we extend the formalism developed in \refref{Friedlander:2022ovf}, and show that there is a viable parameter space where relic decaying particles \cite{Nakayama:2022jza, Carenza:2023qxh, Korochkin:2019pzr, Korochkin:2019qpe, Bolliet:2020ofj, Wadekar:2021qae, Bernal:2022xyi} could lead to direct collapse. Specifically, these could be axion-like particles (\ALPs)~\cite{Peccei:1977hh,Weinberg:1977ma,Wilczek:1977pj,Kim:1979if,Shifman:1979if,Dine:1981rt,Zhitnitsky:1980tq,Svrcek:2006yi,Arvanitaki:2009fg,Acharya:2010zx,Dine:2010cr,AxionLimits} which comprise all of the DM, or it could be a generic particle decaying at high redshifts, for example, one of the many particles in a string axiverse~\cite{Arvanitaki:2009fg}. We model the direct collapse by a self-consistent dynamical evolution model of the DM halo's adiabatic contraction and chemical evolution, which is highly coupled and features significant feedback. On top of this, we reexamine the effect of the baryons' self-shielding of the LW radiation, arguing that the in-situ production of radiation should significantly (if not completely) suppress the effect of shielding. When this is taken into account, we find that this mechanism of direct collapse is indeed viable for fractions of decaying dark matter well below observational constraints~\cite{Poulin:2016nat,Xiao:2019ccl,Alvi:2022aam}. \\

\noindent\textit{Cloud collapse.} The direct collapse of a baryon cloud is a complex dynamical process and hydrodynamic simulation is often needed to study the evolution of the system. However, it has been demonstrated that a simpler approach, the one-zone model \cite{Omukai:2000ic}, can capture the essential ingredients of the direct collapse and offer accurate estimations of key quantities such as the photodissociation rate $k_{\htwo}$ and photodetachment rate $k_{\hminus}$ \cite{Shang:2009ij, Wolcott-Green:2016grm, agarwal2016new}. In the traditional one-zone model, the baryon cloud collapses inside a virialized DM halo that remains constant after the initial top-hat collapse phase. This approximation is acceptable for tracking the temperature and chemical evolution of the cloud only when the LW radiation is treated as a \textit{constant} background. 

The density evolution of the cloud must be supplemented with Boltzmann equations which track $\htwo$ and other chemical components during the collapse. In particular, the dominant $\htwo$ formation channel requires an intermediate product $\hminus$:
\begin{align}
     \mathrm{H}+e^{-} & \rightarrow \mathrm{H}^{-}+\gamma, \\ \mathrm{H}+\mathrm{H}^{-} & \rightarrow \mathrm{H}_2+e^{-}.
\end{align}
The destruction of $\htwo$ can be accomplished by either directly dissociating $\htwo$ with LW photons (photodissociation), or suppression of the formation of $\hminus$ via photons of energy $\gtrsim 0.76$ eV (photodetachment)~\cite{shapiro1987hydrogen}. The success of direct collapse to SMBHs critically depends on the specific radiation intensity $J(E)$ in these energy ranges and consequently the reaction rates $k_{\htwo}$ and $k_{\hminus}$. 

In our model, the LW radiation is directly coupled to the cloud evolution since it comes from the DM halo itself, and it is no longer a constant background. To calculate the DM density during this phase, we adopt a modified one-zone model which is supplemented with an explicit modelling of the adiabatic contraction of the DM halo~\cite{Steigman:1978wqb, Zeldovich:1980st, Ryden:1987ska,blumenthal1986contraction,Eggen:1962dj,Freese:2015mta,Lu:2023xoi}, where the DM halo contracts in response to the collapse of the baryonic cloud. We follow the notations and conventions used in \refref{Lu:2023xoi} with the following exceptions: the photodetachment rate $k_{\gamma}$ in \refref{Lu:2023xoi} only comes from the cosmic microwave background (CMB) spectrum, whereas we replace it here with the rate $k_{\gamma} + k_{\hminus}$. Consequently, this leads to the modification of the equilibrium $\hminus$ fraction: 
\begin{equation}
    x_{\hminus} = \frac{k_{9} x_{\h} x_e n}{k_{\gamma} + k_{\hminus} +(k_{13} + k_{19}) x_e n + (k_{10} + k_{20}) x_{\h} n}.
\end{equation}
In addition, we now must include the photodissociation rate from the LW radiation in the $x_{\htwo}$ equation:
\begin{equation}
    \frac{d x_{\htwo}}{dt} = k_{10} x_{\h} x_{\hminus} n - k_{15} x_{\h} x_{\htwo} n - k_{18} x_e x_{\htwo} n - k_{\htwo} x_{\htwo}.
    \label{xh2}
\end{equation} \\

\noindent\textit{Particle decay.} Let us consider two particle models: a slowly decaying DM particle (as in the \ALP~scenario), and a subdominant component of an extended dark sector which is more rapidly decaying at the epoch of interest. We will refer to the former as DM and the latter as particle $X$. The decay of the particle produces a time dependent radiation-specific intensity $J(E, z)$ (in units of $\mathrm{J/cm^2/s/Hz/sr}$) in the gas cloud given by \cite{Friedlander:2022ovf},
\begin{equation}
    J(\vec{r}, E, z)=\frac{E}{4 \pi} \int d V^{\prime} \frac{d n_\gamma}{d E d t}(\vec{r^{\prime}}, E, z) \frac{1}{(\vec{r^{\prime}}-\vec{r})^2}.
\end{equation}
This expresssion---although completely general---is computationally intensive since keeping track of the spatial dependence is highly impractical. In addition, we take the DM halo and the baryon cloud to have uniform density in the one-zone model, so the reaction rates are also position independent. This motivates us to simplify the specific intensity by using a spherically symmetric halo and computing its value at the center:
\begin{equation}
    J(E, z)=E \int d r^{\prime} \frac{d n_\gamma}{d E d t}\left(r^{\prime}, E, z\right)~.
\end{equation}
The one-zone differential injection rate, $d n_{\gamma}/d E d t$ is given by
\begin{equation}
    \frac{d n_\gamma}{d E d t}(E, z)=\frac{f_X(z) \Gamma \bar{\rho}_{DM}}{m }\frac{d N}{d E}(E)~,
    \label{diff_inj}
\end{equation}
where $f_X$ is the time dependent energy fraction of $X$ compared to the total DM (if our decaying particle is all of the DM, $f_X=1$). In \equaref{diff_inj}, we assume that photons are the only decay products and neglect other possibilities, such as neutrinos or dark radiation (if they were to be included, our results will simply scale with the branching ratio of the photon decay channel). Our one-zone approximation contrasts with the `critical curve' approach where $J(E, z)$ is calculated using a non-trivial halo density profile \cite{Friedlander:2022ovf}. We will map our results to the critical curve plot for comparison.

The shape of the decay spectrum $dN / dE$ depends on the number of decay products. For \ALPs, it is well-motivated to consider two body decay:
\begin{equation}
    E \frac{dN}{dE} = 2 \delta\left(1 - \frac{2E}{m_{X}}\right)~.
\end{equation}
Following \refref{Friedlander:2022ovf}, we also consider three body decay with an energy-independent decay amplitude, modelled with the so-called `parabola spectrum':
\begin{equation}
    E \frac{d N}{d E}=6\left[\frac{E}{m_{X}}-\left(\frac{E}{m_{X}}\right)^2\right] \Theta\left(1-\frac{E}{m_{X}}\right).
\end{equation}

The chemical rates $k_{\hminus}$ and $k_{\htwo}$ can be computed using the specific intensity \cite{Wolcott-Green:2016grm}:
\begin{align}
    k_{\mathrm{H}^{-}}(z)&=\int_{0.76 \mathrm{eV}}^{13.6 \mathrm{eV}} 4 \pi \sigma_{\mathrm{H}^{-}}(E) \frac{J(E, z)}{E} \frac{d E}{h}~\nonumber\\
    k_{\mathrm{H}_2}(z) &\approx 1.39 \times 10^{-12} \mathrm{~s}^{-1}\frac{J_{LW}(z)}{J_{21}},
    \label{kh2}
\end{align}
where we take the cross section $\sigma_{\mathrm{H}^{-}}$ given in \refref{shapiro1987hydrogen} and $J_{21} = 10^{-21} \mathrm{erg} \mathrm{~s}^{-1} \mathrm{~Hz}^{-1} \mathrm{~cm}^{-2} \mathrm{~sr}^{-1}$. Here, $J_{LW}$ is calculated by taking the average of $J(E, z)$ in the LW band due to the complex structure of the rotational-vibrational states of $\htwo$.  

We note regarding \equaref{kh2} that previous studies used either constant spectra or black body spectra with temperatures $10^4~\rm K$ or $10^5~\rm K$ \cite{Omukai:2000ic, Shang:2009ij, Hartwig:2015ura, Inayoshi:2014nha, Latif:2014cda}. While the three body decay spectrum is still continuous and will not deviate far from the approximations in \equaref{kh2}, two body decay could potentially alter the dissociation rate, as photons from the decay are monochromatic, and the absorption spectrum of $\htwo$ is discrete.  This does not present a problem: the typical timescale for the collapse is $\Delta z \sim 1$, during which the photon energy is redshifted by $\sim 0.5 ~\rm eV$ if the halos collapse around $z = 20$. The redshift and the thermal Doppler broadening allow the radiation to cover the fine spaced rotational-vibrational energy levels~\cite{morton1976interstellar}. The spectral energy distribution (SED) in our model is different from the black body SED widely studied in the literature. This fact, together with the dynamic nature of the radiation intensity, makes a direct comparison with the critical intensity found in previous simulations impractical.\\

\noindent\textit{Shielding.} The rate given in \equaref{kh2} is valid only in the optically thin regime, and additional treatment is necessary when the column density of $\htwo$ reaches the critical value of $ 10^{14} ~\rm cm^{-2}$. Beyond this density, the $\htwo$ becomes optically thick to LW radiation and the gas is self-shielding, reducing the dissociation rate by a fraction $f_{\rm shield}(N_{\htwo}, T)$. We adopt a shielding fraction as in \refref{Draine:1996hna} that takes into account thermal broadening effects. The full density dependent dissociation rate is then
\begin{equation}
    k_{\htwo}(N_{\htwo}, T) = k_{\htwo} f_{\rm shield}(N_{\htwo}, T).
    \label{fullkh2}
\end{equation}
In practice, computing the $\htwo$ column density is intractable in cosmological simulations. In Ref. \cite{wolcott2011photodissociation}, it was shown that the column density is best approximated by 
\begin{equation}
    N_{\htwo} = \frac{1}{4} n_{\htwo} \lambda_{\rm Jeans},
\end{equation}
where $\lambda_{\rm Jeans}$ is the Jeans length. We will use this equation to calculate the column density and the shield fraction.

The magnitude of $f_{\rm shield}$ is a crucial bottleneck in previous studies of the LW radiation required for direct collapse~\cite{wolcott2011photodissociation, Wolcott-Green:2016grm, Shang:2009ij, luo2020direct} since in the case where the LW flux is anisotropic and sourced exterior to the cloud, the success of direct collapse to SMBHs crucially depends on the ability of the radiation to penetrate the outer shell of molecular hydrogen to reach the core region. Our scenario different: the baryon cloud is immersed in the nearly homogeneous and isotropic background of LW photons from the relic particle decay, and the core region with the highest DM density experiences the strongest radiation intensity. Therefore,  the LW radiation can reach the $\htwo$ molecules in the core region even in the optically thick regime. Since the radiation no longer needs to penetrate through the cloud, the shield fraction is greatly reduced. Furthermore, as the shielding is reduced, the LW photons can dissociate $\htwo$ more effectively, further decreasing the shielding. In addition, since the DM halo is larger than the baryonic cloud, the exterior region of the baryonic cloud is irradiated by the exterior portion of the halo as well as in-situ photons. Even if the DM is less dense in this region, the significant exterior flux makes it unlikely that the cloud could fragment in the exterior regions but not the interior, as was suggested in Ref.~\cite{Friedlander:2022ovf}.

To quantify the uncertainty in the shielding, we introduce a new parameter $\varepsilon_{sh}$ to track how much radiation is being shielded. We parametrize the in-situ shield fraction by
\begin{equation}
    f_{\rm in-situ} = 1 - \esh (1 - f_{\rm shield}),
\end{equation}
so that $\esh = 1$ corresponds to full self-shielding and $\esh = 0$ corresponds to no shielding. We present our results for several values of this shielding paramater, although we expect that it should be very close to $0$.\\

\noindent\textit{Particle decay constraints.} For both these mechanisms, we must tune the mass of the decaying particle so that they produce LW photons. In addition, the fraction of decaying dark matter must satisfy observational constraints~\cite{Poulin:2016nat,Xiao:2019ccl,Alvi:2022aam} which require no more than $\sim1\%$ of dark matter to decay in the early universe, although as we will see the decay fractions we require here for a successful collapse fall easily shy of this mark.

A natural candidate for decaying DM in the LW mass range are \ALPs~\cite{Peccei:1977hh,Weinberg:1977ma,Wilczek:1977pj,Kim:1979if,Shifman:1979if,Dine:1981rt,Zhitnitsky:1980tq,Svrcek:2006yi,Arvanitaki:2009fg,Acharya:2010zx,Dine:2010cr,ohare_cajohareaxionlimits_2020,OHare:2024nmr} since they are well-motivated dark matter candidates which naturally decay to two photons. In this scenario, we must restrict the mass to range from $22.4$ eV to $27.2$ eV. Although this region cannot contain the QCD axion~\cite{ohare_cajohareaxionlimits_2020,OHare:2024nmr}, the parameter space is still available for the more generic \ALPs, with the strongest constraints from cosmic optical background (COB) observations~\cite{Nakayama:2022jza, Carenza:2023qxh}. 

\begin{figure}[!ht]
    \includegraphics[width=\columnwidth]{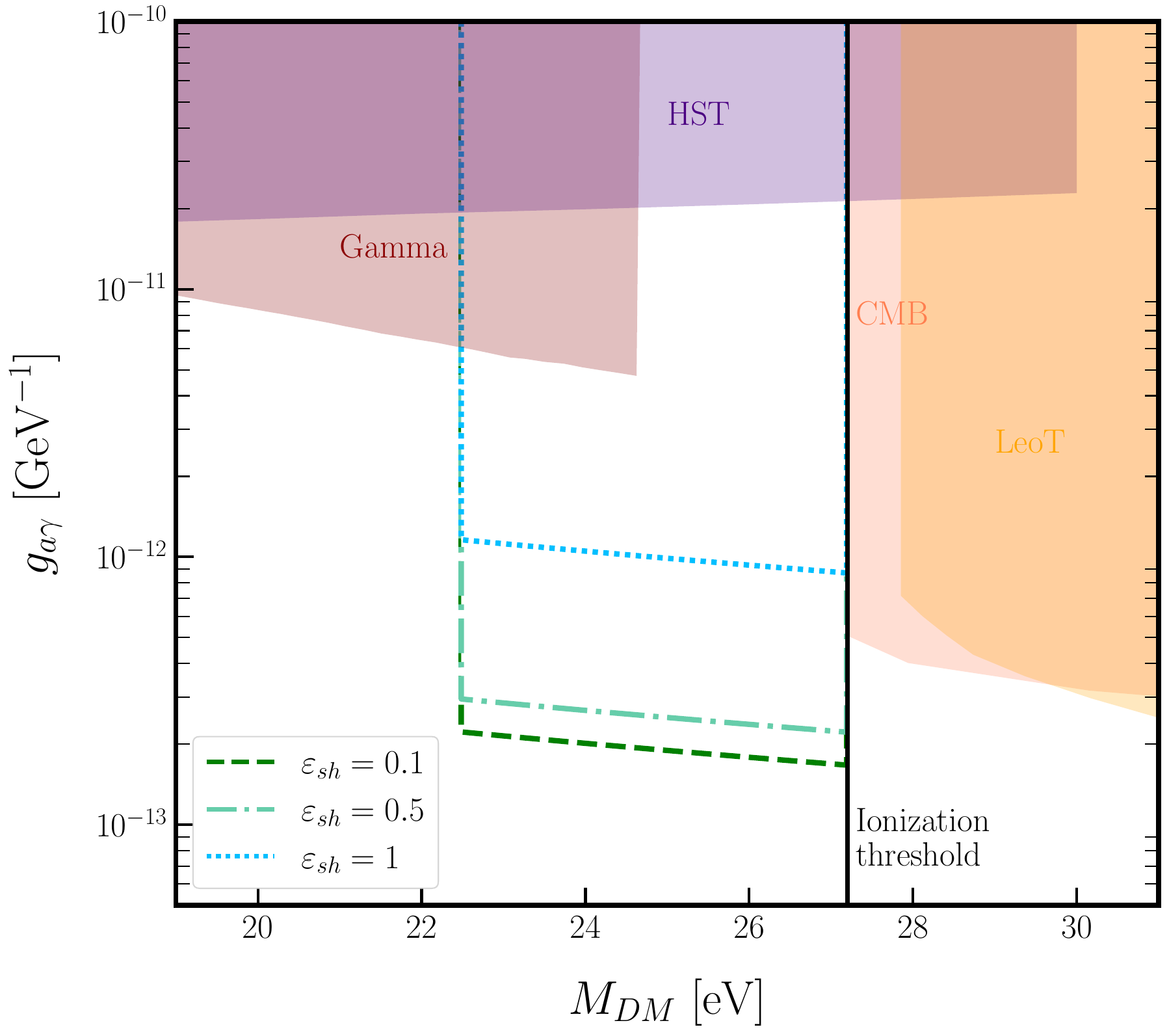}
    \caption{ \ALP~parameter space and relevant constraints~\cite{ohare_cajohareaxionlimits_2020,OHare:2024nmr,Nakayama:2022jza, Carenza:2023qxh,Korochkin:2019pzr, Korochkin:2019qpe, Bernal:2022xyi,Bolliet:2020ofj,Acharya:2023bln,Wadekar:2021qae} for successful direct collapse at $z\sim 20$. We demonstrate the results for three choices of $\esh$.}
    \label{fig:alp_para}
\end{figure}

The parameter space in the conventional axion-photon coupling $g_{a \gamma}$ and dark matter mass $M_{DM}$ is shown in \figref{fig:alp_para}, where the minimum decay rates for the successful collapse are shown for different shielding assumptions and will be discussed in more detail in the following section.
We include the constraints from the Hubble Space Telescope (HST) observation of cosmic optical background (COB) anisotropy \cite{Nakayama:2022jza, Carenza:2023qxh} and gamma-ray attenuation \cite{Korochkin:2019pzr, Korochkin:2019qpe, Bernal:2022xyi}. In addition, if the \ALP~mass exceeds the hydrogen ionization threshold, it is strongly constrained by CMB anisotropies \cite{Bolliet:2020ofj,Acharya:2023bln} and the heating of the dwarf galaxy Leo T \cite{Wadekar:2021qae}. Notably, in Ref. \cite{Bolliet:2020ofj}, the constraint is derived assuming a photon injection spectrum with a narrow but finite width, leading to constraints that saturate below the ionization threshold at $27.2~\rm eV$. This assumption is not valid in the case of \ALP~decay, so we cut off this constraint at the ionization threshold. We emphasize that the available parameter space for $\esh=1$ is not eliminated even if we use the CMB anisotropy constraint without the cutoff. 

For the more generic short-lived particle, a well-motivated scenario is the {\it string axiverse}~\cite{Arvanitaki:2009fg}, in which there are naturally many BSM particles over a wide range of masses.  Just one of these particles would need to have the appropriate mass to accommodate our proposed mechanism. In this scenario, the relevant constraints are those which constrain the maximum evaporating fraction of dark matter~\cite{Bolliet:2020ofj,Poulin:2016nat,Xiao:2019ccl,Alvi:2022aam}. 
\\

\begin{figure}[ht!]
    \begin{subfigure}{0.9\columnwidth}
        \includegraphics[width=\textwidth]{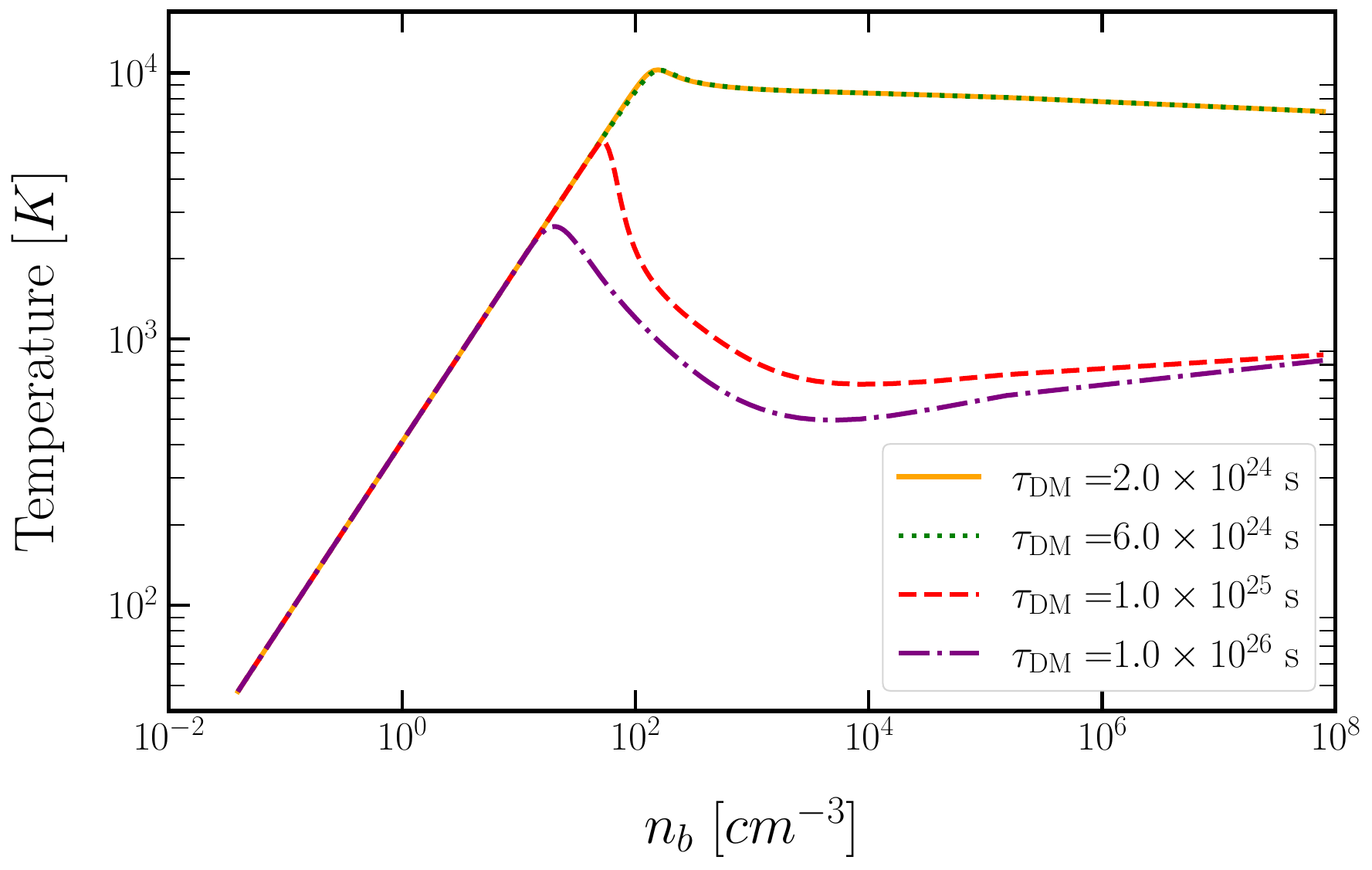}
    \end{subfigure}
    \begin{subfigure}{0.9\columnwidth}
        \includegraphics[width=\textwidth]{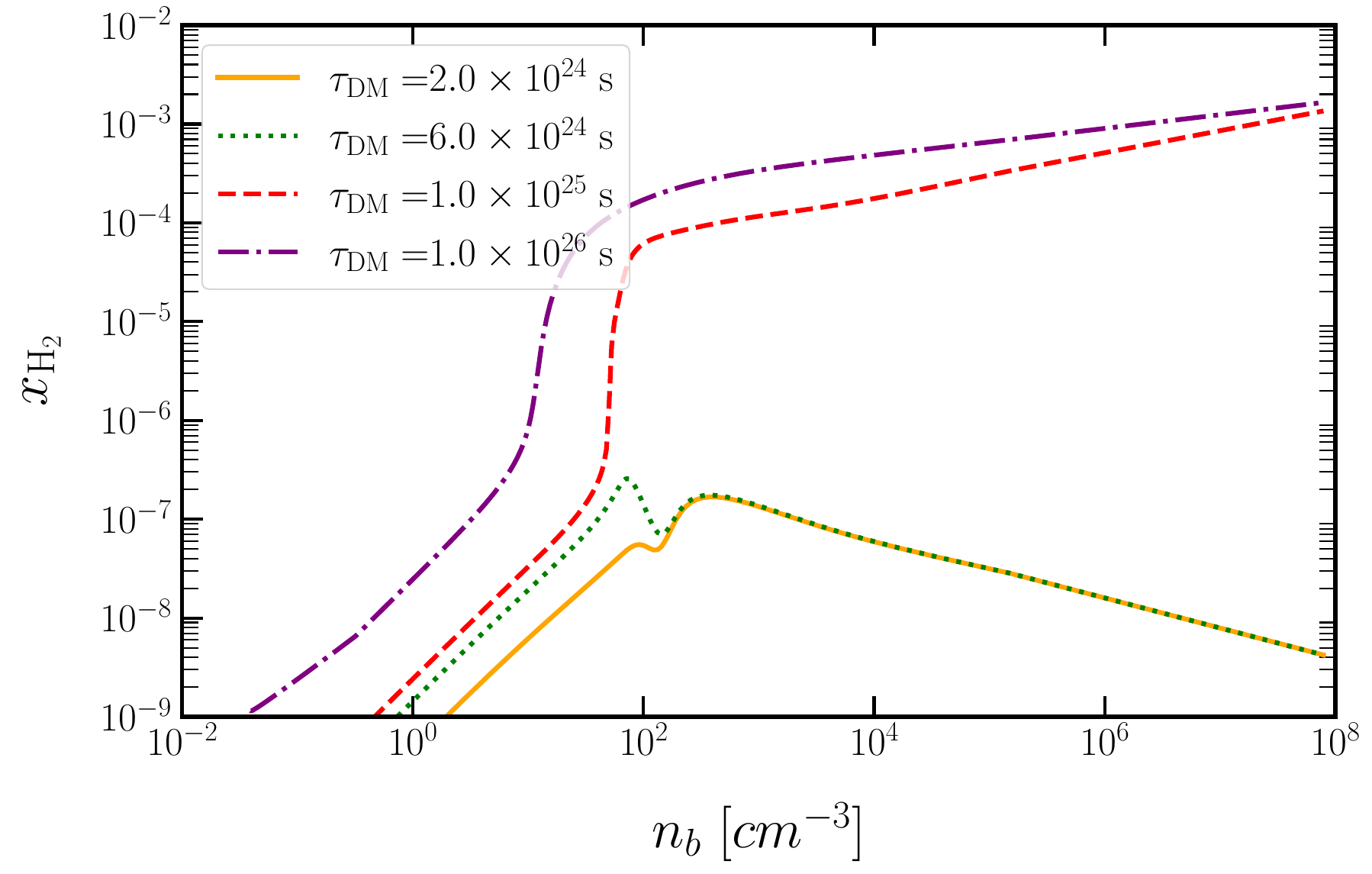}
    \end{subfigure}
    \captionsetup{width=\columnwidth}
    \caption{Temperature and $\htwo$ fraction during the collapse in the case of DM undergoing two body decay. We see a clear bifurcation behavior here: if the DM lifetime is below a threshold value $\sim 6\times 10^{24}$ s, the formation of $\htwo$ is inhibited and consequently, the temperature stays at $10^4$ K during collapse. The DM mass is chosen to be 25 eV and the halo collapses at $z\sim 20$. }
    \label{fig:evolution}
\end{figure}

\noindent\textit{Results and discussion.} 
An example of the temperature and chemical evolution during the collapse is shown in \figref{fig:evolution}, where one can understand the forward progress of time as moving to the right on the curves. We observe a bifurcation behavior between the successful and failed collapse process: a successful direct collapse is characterized by a final temperature near $10^4$ K with $\htwo$ fraction $\ll 10^{-5}$. These criteria are used to check the collapse outcome in our parameter space search.

\begin{figure}[!ht]
    \centering
    \captionsetup{width=\columnwidth}
    \includegraphics[width=\columnwidth]{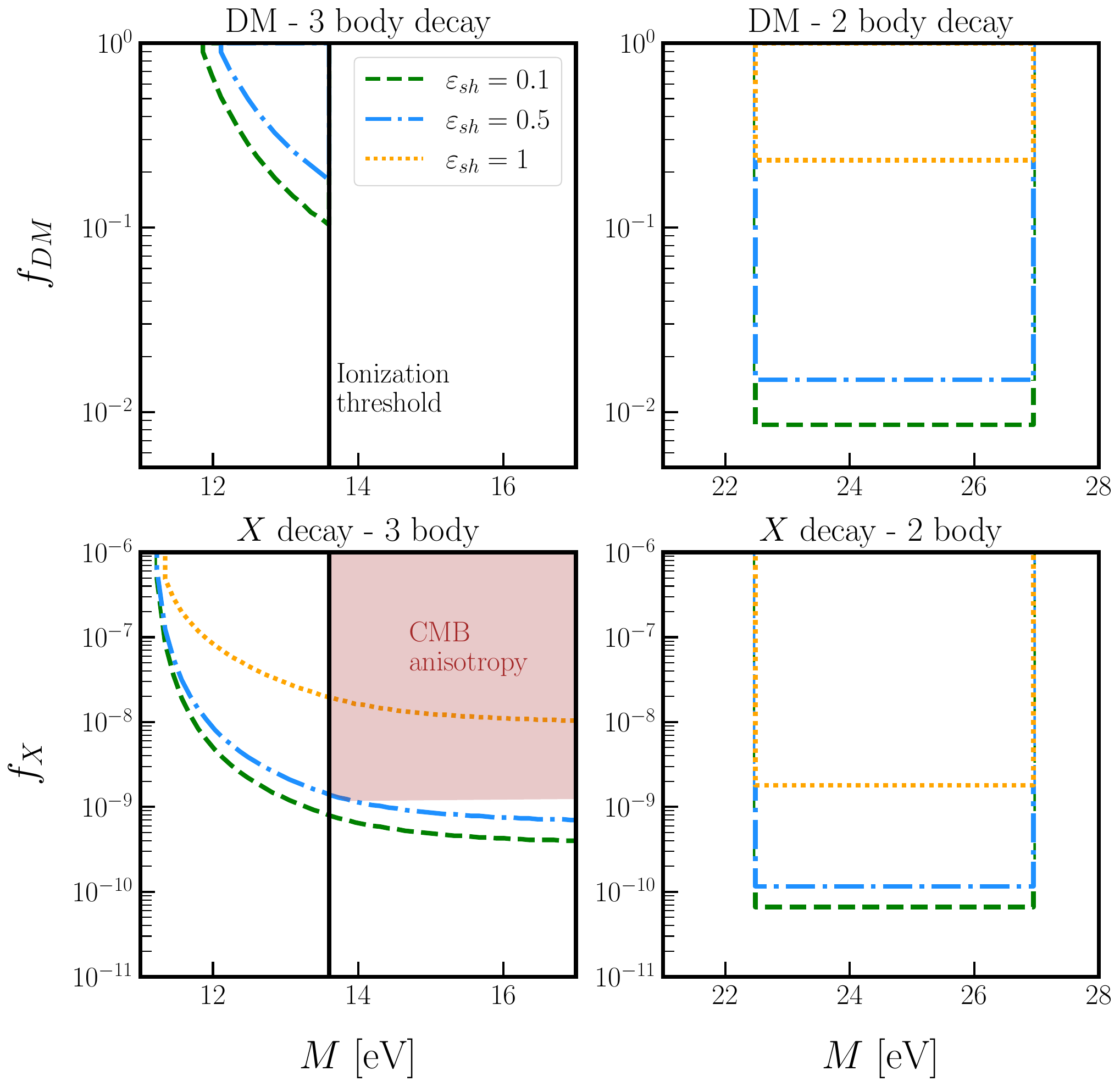}
    \caption{Energy fractions of particle decay required to trigger direct collapse in the case of early decaying X and DM, where the three curves on each plot are for different amounts of shielding. We show the parameter space for three body decay (left panels) and two body decay (right panels). The DM life time is chosen to be $2\times 10^{24}$ s and X is set to decay at $z=22$. The only relevant constraint is from the CMB anisotropy for X decay~\cite{Bolliet:2020ofj}, which is plotted on the bottom left panel. The halo is set to collapse at $z\sim 20$ here.}
    \label{fig:f_m_para}
\end{figure}

In the case of the $X$ particle decaying early (around $z\sim 20$ here), only a small fraction of the total DM density is required to allow direct-collapse SMBH formation. We show the viability of this scenario, together with the DM decay, in \figref{fig:f_m_para}. As seen from the top left panel of \figref{fig:f_m_para}, if self shielding is not reduced, DM with three body decay cannot produce enough LW radiation for direct collapse. This is consistent with the critical curve results obtained in \refref{Friedlander:2022ovf}. For photons below the hydrogen ionization threshold, the fraction $f_{X}$ (at a time long before the epoch of decay) is constrained to be $\lesssim 10^{-2}$ from CMB spectral distortion observed by COBE/FIRAS. Above the threshold, however, the injection is subject to strong constraints from CMB anisotropies, and we include this constraint on the bottom left panel for a particle with lifetime $10^{14}$~s. For comparison, we are interested in lifetimes around $10^{15}$~s, reducing the plotted constraints. We do not need to tune the particle lifetime to exactly coincide with the collapse time. For a halo to collapse at $z=20$, viable choices for the $X$ lifetime span a wide range from $z\sim 50$ to $z\sim 4$, because decay is a stochastic process. 

\begin{figure}[ht!]
    \centering
    \captionsetup{width=\columnwidth}
    \includegraphics[width=0.9\columnwidth]{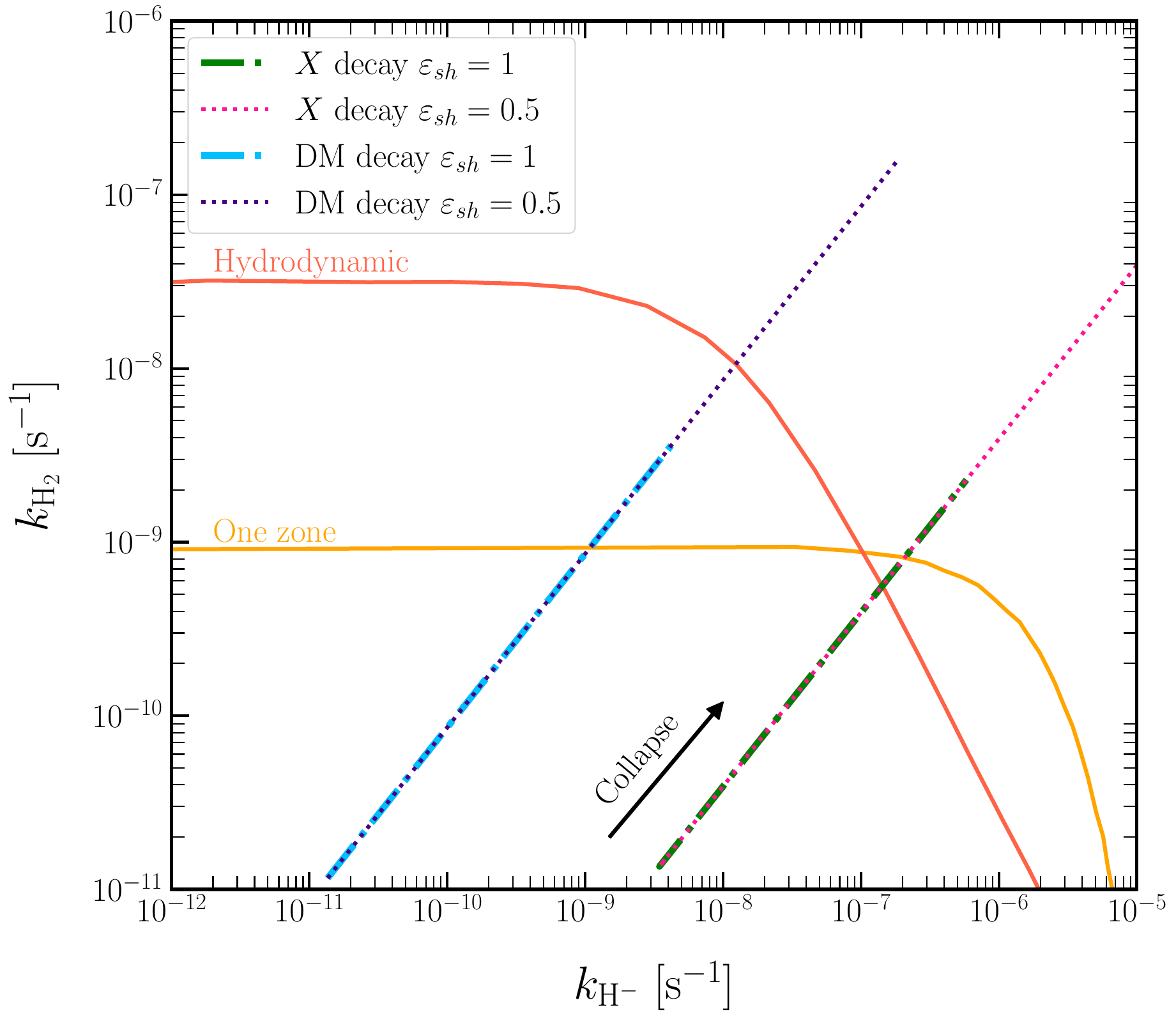}
    \caption{Comparison of the critical curves with the reaction rates (in a successful direct collapse) calculated in our dynamical one-zone approach. The critical curves are taken from \refref{Wolcott-Green:2016grm} (one-zone model) and \refref{luo2020direct} (hydrodynamic simulation). Both critical curves take $\lambda_{\rm Jeans} / 4$ as the characteristic length for computing the $\htwo$ column density. For a moderate reduction of the shield fraction, the final rates go beyond the critical values required even in hydrodynamic simulations. The rates are calculated for an early decayed particle (three body decay) with mass $13$ eV and $f_X = 10^{-7}$ that decays at $z=25$, and for DM (two body decay) with mass $25$ eV and lifetime $2\times 10^{24}$ s.}
    \label{fig:critical}
\end{figure}

When the LW radiation is constant, it is convenient to derive the critical dissociation and detachment rates required as a simple check for successful direct collapse. Such requirement, plotted on the $k_{\htwo}-k_{\hminus}$ plane, forms the so called critical curve~\cite{agarwal2016new, Wolcott-Green:2016grm, luo2020direct}. However, a direct comparison between our results and the critical curve is not straightforward. Firstly, the LW produced by particle decay is no longer a single point on the $k_{\htwo}-k_{\hminus}$ plane, with lower reaction rates (due to lower specific intensity) at the beginning of the collapse but increasing significantly at later stage. Another obstruction comes from the fact that all previous critical curve results \cite{agarwal2016new, Wolcott-Green:2016grm, luo2020direct} assumed an external source of LW radiation with full self-shielding ($\esh = 1$). In our case, the shield fraction is expected to be heavily reduced due to the in-situ emission of LW photons. Nevertheless, we present the critical curves in \figref{fig:critical} as a useful check. Naively it would seem that the LW radiation cannot reach the critical value required from hydrodynamic simulations when $\esh = 1$, but even a moderate reduction of the shield fraction can increase the final rates by several orders of magnitude, exceeding the threshold required for successful collapse.

\noindent\textit{Conclusion.} 
We have shown that the decay of DM or a short-lived particle $X$ can effectively halt the production of molecular hydrogen, allowing direct collapse to a SMBH to occur. While previous studies found that particle decay may not lead to collapse, we found that dynamically including the adiabatic contraction of the DM halo, as well as considering lower shielding than in the case where the LW background comes from an exterior source, allowed our model to be successful. To go beyond the one-zone approach that we adopt in this work, a full-scale simulation that takes into account a realistic halo profile and in-situ effects---similar to the one in~\refref{Stacy:2013xwa}---is needed to confirm our result.

The decay of \ALPs~in our mass range of interest can produce light in the optical and UV range. Interestingly, as  pointed out in \refref{Bernal:2022wsu}, such decay could explain the COB excess observed by New Horizons’ Long Range Reconnaisance Imager (\textsc{Lorri}) \cite{Lauer:2022fgc}. Later studies, however, concluded that \ALP~decay is unlikely to produce such an excess due to the COB anisotropy at $606$ nm and gamma ray attenuation \cite{Nakayama:2022jza, Carenza:2023qxh, Wang:2023imi, Bernal:2022xyi}. Probing these so called `blue axions' offers an opportunity to test our proposed mechanism. With future HST measurements at higher frequencies, a large portion of our parameter space could be explored~\cite{Carenza:2023qxh}. It would be of great interest if the mystery of high redshift SMBHs and the question of DM can be answered simultaneously.

\section*{Acknowledgements}
\noindent 
This work was supported by the U.S. Department of Energy (DOE) Grant No. DE-SC0009937. The work of A.K. was also supported by World Premier International Research Center Initiative (WPI), MEXT, Japan, and by Japan Society for the Promotion of Science (JSPS) KAKENHI Grant No. JP20H05853. 
This work made use of N\textsc{um}P\textsc{y}~\cite{numpy2020Natur.585..357H}, S\textsc{ci}P\textsc{y}~\cite{scipy2020NatMe..17..261V}, and M\textsc{atplotlib}~\cite{mpl4160265}.

\bibliographystyle{bibi}

\bibliography{main.bib}

\end{document}